\documentclass[11pt,a4,cite]{article}
\usepackage{epsfig}
\usepackage{graphicx}
\def\ga{\alpha}
\def\gb{\beta}

\def\ge{\epsilon}
\def\gg{\gamma}
\def\gd{\delta}

\def\gm{\mu}

\def\gp{\pi}

\def\gt{\theta}

\def\delp{\partial_+}
\def\delm{\partial_-}

\def\part{\partial}
\def\hlf{\frac{1}{2}}
\def\A0{A^{+}_0}

\def\xpl{x^{+}}
\def\xmin{x^{-}}
\def\ymin{y^{-}}
\newcommand{\nc}{\newcommand}
\nc{\intgl}{\int\limits_{-L}^{+L}\!{{dx^-}\over\!2}}
\nc{\intgly}{\int\limits_{-L}^{+L}\!{{dy^-}\over\!2}}
\nc{\zmint}{\int\limits_{-L}^{+L}\!{{dx^-}\over{\!2L}}}
\def\beq{\begin{equation}}
\def\eeq{\end{equation}}
\def\bea{\begin{eqnarray}}
\def\eea{\end{eqnarray}}
\begin{document}
\title{\bf Analytic solution of the microcausality problem in Discretized Light 
Cone Quantization}
\vspace{50pt}
\author{
\phantom{abcd}\\
\phantom{abcd}\\
L$\!\!$'ubom\'{\i}r Martinovi\v c\\ 
{\it Institute of Physics, Slovak Academy of Sciences} \\
{\it D\'ubravsk\'a cesta 9, 845 11 Bratislava, Slovakia} 
\thanks{\tt e-mail: fyziluma@savba.sk} \\ 
\vspace{0.1mm}\\
and\\
\vspace{0.1mm}\\
Marshall Luban \\
{\it Ames Laboratory and Department of Physics and Astronomy,}\\
{\it Iowa State University, Ames, Iowa 50011,USA}
\thanks{\tt e-mail: luban@ameslab.gov}}
\date{}
\maketitle
\begin{abstract}

It is shown that that violation of causality in two-dimensional light-front 
field theories quantized in a finite ``volume'' $L$ with periodic or 
antiperiodic  boundary conditions is marginal and vanishes smoothly 
in the continuum limit. For this purpose, we derive an exact integral 
representation for the complete infinite series expansion of the two-point 
functions of a free massive scalar and fermi field for an arbitrary finite 
value of $L$ and show that in the $L \rightarrow \infty$ limit we retrieve 
the correct continuum results.
 
\end{abstract}
 
\newpage

\section{Introduction}

Light-front field theory has some important advantages \cite{Dir,LKS,Rohr}
due to it simplified vacuum structure that make it a promising
theoretical framework for elementary particle physics.  On the other hand, a
systematic understanding of non-perturbative structure of light-front (LF)
field theory has so far not been achieved. A particularly advantageous
formulation appears to be LF quantization in a finite-volume, known as  
discretized light-cone quantization (DLCQ) \cite{MY,PB}. It incorporates 
in an efficient way boundary conditions which are required for mathematical 
consistency even in the continuum formulation \cite{Steinh}. The point is 
that the numerous LF constraints which reduce
the number of independent field degrees of freedom can be (at least in
principle) uniquely inverted only if the corresponding Green's functions
satisfy (anti)periodic boundary conditions. For periodic boundary conditions,
one may then study physical implications of zero-mode operators (carrying the 
LF momentum $p^+=0$) in a finite volume which serves as infrared 
regularization. Thus, it is desirable to formulate light-front quantization 
in a finite space volume with a discrete infinity of modes as a quantum field 
theory in its own right. This implies that one should verify all usual 
well-established properties (such as causality, Poincar\'e invariance, 
singularity structure, etc.) in this framework to check its overall consistency.

It is far from clear that these properties will hold true. Indeed, already
for the simple question of microcausality of a massive scalar field in two
dimensions it has been concluded that causality is violated by the infrared
finite-volume regularization \cite{HKS}. More precisely, it has been argued
that periodic boundary conditions are incompatible with causal behavior
of the light-front quantum theory. The method used to demonstrate this was 
basically a numerical study of the corresponding series truncated at some 
value of discretized LF momentum $p^+$ for which the results stabilized. This 
method gave a very satisfactory picture of the causality in a space-like box, 
namely vanishing (up to negligible numerical effects) of the Pauli-Jordan (PJ) 
function (which is twice the imaginary part of the full Wightman function)  
for space-like separations and a usual oscillatory behaviour in the 
time-like region. This picture however failed completely for a LF system 
restricted to $-L \leq \xmin \leq L$. Not only did the numerical results for 
the PJ function fail to vanish for $x^2 < 0$, but it was even found not to 
converge to the correct continuum expression. Two obvious explanations are: 
1. the discretized light-front theory has some fundamental difficulty or, 
2. the results of the numerical computations are misleading or at least do 
not reveal the full nature of the problem. To clarify this issue it would be 
preferable to find a method for analytical evaluation of the infinite series 
expansion of the PJ function, corresponding to the integral representation 
of the PJ function in the continuum formulation. 

A calculation of this kind has been sketched in Ref.\cite{GSW}
indicating that the PJ function computed in a finite volume converges
to the correct continuum form in the large $L$ limit. The results of  
Ref.\cite{HKS}, however, put this picture
into doubt (see also Ref.\cite{Haricau} for a careful numerical 
analysis). We find it very important to clarify the situation by an
independent analytical calculation and this is the main purpose of the present
paper. More generally, we wish to show that there is a natural mathematical
formalism for evaluation of infinite series corresponding to various
correlation functions of the discretized LF theory. This formalism is very
well adapted to the form of LF kinematics and dispersion relation and it uses
some properties of polylogarithm functions \cite{Lew}. As a result, an integral
representation can be given for the infinite series expansion,
in particular for that corresponding to the free-field Wightman functions. This
representation explicitly depends on the box length $L$. We use analytical 
methods to study the large $L$ dependence of the integral representation 
of the PJ function. The result of our analysis is unambiguous: there is no 
physically relevant violation of causality. We recover the well known 
continuum result plus finite-volume corrections which are suppressed by a 
power of $L$ and thus vanish in the large-$L$ limit. Moreover, even for 
relatively small value of $L$ one obtains a very plausible picture: an 
oscillatory structure in the time-like region and near-vanishing values of 
the PJ function in the space-like region. This will be demonstrated 
numerically in Sec.4 below.

\section{Free-field correlation functions}

We first briefly describe the derivation of the two-point Wightman
function for a massive LF scalar field in two dimensions. In the continuum
formulation, the mode expansion of the scalar field is
\footnote{We use the convention $x=(\xpl,\xmin), x^2=\xpl\xmin, 
\partial^{\mp}=2\partial_{\pm}= \partial / \partial x^{\pm}$}
\beq
\phi(x) = \int\limits_0^{\infty}{dq^+ \over {4\gp q^+}}
\big[a(q^+,\xpl)e^{-{i \over 2}q^+\xmin} +
a^\dagger(q^+,\xpl)e^{{i \over 2}q^+ \xmin}\big].
\label{scalexp}
\eeq
In the free case, the time evolution is simple. From the Klein-Gordon equation
we have $a(q^+,\xpl)=a(q^+,0)\exp(-i\hat{p}^- \xpl/2),\;
\hat{p}^- = \gm^2/p^+$, and hence
\beq
\phi(x) = \int\limits_0^{\infty}{dq^+ \over {4\gp q^+}}
\big[a(q^+)e^{-{i \over 2}q^+\xmin -{i \over 2}{\gm^2 \over q^+}\xpl} +
a^\dagger(q^+)e^{{i \over 2}q^+ \xmin + {i \over 2}{\gm^2 \over q^+}\xpl}\big],
\label{frexp}
\eeq
where $a(q^+) \equiv a(q^+,0)$ and small imaginary parts for $x^\pm$ are
understood (see below). The corresponding two-point function 
\beq
D(x) = \langle 0 \vert \phi(0)\phi(x) \vert 0 \rangle =
\int\limits_0^{\infty}{dp^+ \over 4\gp p^+}
e^{{i \over 2}p^+\xmin  +  {i \over 2}{\gm^2 \over p^+}\xpl}
\label{D}
\eeq
can easily be obtained using the commutation relation
\beq
\big[a(p^+),a^\dagger(p^{\prime +}\big]=4\gp p^+\gd(p^+-p^{\prime +}).
\label{Fcr}
\eeq
With the help of known integral formulae 3.471 from \cite{GR}, one further finds
\bea
D(x) &=& -\frac{1}{8}\vert {\ge} (\xpl) + {\ge} (\xmin) \vert
\Big(N_0(\gm\sqrt{x^2}) - i~ {\ge}(\xpl) J_0(\gm\sqrt{x^2})\Big)  
\nonumber \\
&+& \frac{1}{4\gp} \vert {\ge} (\xpl) - {\ge} (\xmin) \vert K_0(\gm\sqrt{-x^2}),
\label{F1a}
\eea
where $\ge(x^\pm)$ is the sign function.
To guarantee the convergence in equations \ref{D} and \ref{F1} below, it is 
assumed that the quantities $x^\pm$ include an infinitesimal positive 
imaginary part.
$J_0(z), K_0(z)$, and $N_0(z)$ are standard Bessel, modified Bessel, and 
Neumann functions.

The corresponding expression for the same system quantized in a finite
volume with field obeying periodic boundary condition reads
\beq
\hat{D}(x) = \frac{1}{4\gp}\sum_{n=1}^{\infty} \frac{1}{n}e^{\frac{i}{2}
p^+_n\xmin + \frac{i}{2} \frac{\gm^2}{p^+_n}\xpl},\;\;p^+_n=\frac{2\gp}{L}n,
\label{df1}
\eeq
where the field expansion
\beq
\phi(x) = \frac{1}{\sqrt{2L}}\sum_{n=1}^{\infty}\frac{1}{\sqrt{p^+_n}}\Big[
a_n e^{-\frac{i}{2} p^+_n\xmin - \frac{i}{2}\frac{\gm^2}{p^+_n}\xpl} +
a^\dagger_n e^{\frac{i}{2} p^+_n\xmin + \frac{i}{2}\frac{\gm^2}{p^+_n}
\xpl}\Big]
\label{dfexp}
\eeq
with
\beq
\left[a_m,a^\dagger_n\right] = \gd_{m,n}
\label{dcr}
\eeq
has been used. The mode with $n=0$ is excluded in the above series since the
free Klein-Gordon equation $(4\delp\delm + \gm^2)\phi(x)=0$ is not
compatible with a non-vanishing Fouri\'er mode carrying $p^+=0$ if $\gm \neq 0$.

A similar treatment can be given for free massive LF fermions. In the 
representation where the $\gg_5$-matrix is diagonal, the LF Dirac equation for 
the spinor field $\psi^T = (\psi_1, \psi_2)$ of mass $m$ separates into two 
equations for the components $\psi_1, \psi_2$: 
\beq
2i\delp \psi_2(x)=m\psi_1(x),~~2i\delm \psi_1(x)=m\psi_2(x),
~~\gg_5={\rm diag}(-1,1)
\label{Dir}
\eeq
The independent component $\psi_2$ can be expanded at $\xpl=0$ into the Fouri\'er 
integral with the operator coefficients $b(p^+),b^\dagger(p^+),d(p^+),d^\dagger(p^+)$  
and the above dynamical equation determines its LF time evolution as  
\beq
\psi_2(x) = \int\limits_0^{\infty}{dp^+ \over {4\gp \sqrt{p^+}}}
\big[b(p^+)e^{-{i \over 2}p^+\xmin -{i \over 2}{m^2 \over p^+}\xpl} +
d^\dagger(p^+)e^{{i \over 2}p^+ \xmin + {i \over 2}{m^2 \over p^+}\xpl}\big].
\label{psi2}
\eeq
The solution of the constraint equation is 
\beq
\psi_1(x) = m\int\limits_0^{\infty}{dp^+ \over {4\gp \sqrt{p^+}p^+}}
\big[b(p^+)e^{-{i \over 2}p^+\xmin -{i \over 2}{m^2 \over p^+}\xpl} - 
d^\dagger(p^+)e^{{i \over 2}p^+ \xmin + {i \over 2}{m^2 \over p^+}\xpl}\big]. 
\label{psi1}
\eeq
The quantization rule 
\beq
\big\{\psi_2(0,\xmin),\psi_2^\dagger(0,\ymin)\big\} = 
\hlf \gd(\xmin-\ymin)
\label{antic}
\eeq
is equivalent to the following anticommutation relations for the Fock operators:
\beq
\big\{b(p^+),b^\dagger(p^{\prime +})\big\}=\big\{d(p^+),
d^\dagger(p^{\prime +})\big\}= 4\gp p^+ \gd(p^+ - p^{\prime +}). 
\label{Fockanti}
\eeq
It is very simple to calculate the two-point Wightman functions $S_{\ga\gb}(x)$: 
\bea
&&S_{22}(x) = \langle 0 \vert \psi_2(0)\psi^\dagger_2(x) \vert 0 \rangle =
\int\limits_0^{\infty}{dp^+ \over 4\gp}
e^{{i \over 2}p^+\xmin +  {i \over 2}{m^2 \over p^+}
\xpl} ~~~~~~~~~ \nonumber \\
&&S_{11}(x) = \langle 0 \vert \psi_1(0)\psi^\dagger_1(x) \vert 0 \rangle =
m^2\int\limits_0^{\infty}{dp^+ \over 4\gp {p^+}^2}
e^{{i \over 2}p^+\xmin  +  {i \over 2}{m^2 \over p^+}
\xpl } \nonumber \\
&&S_{12}(x) = \langle 0 \vert \psi_1(0)\psi^\dagger_2(x) \vert 0 \rangle =
m \int\limits_0^{\infty}{dp^+ \over 4\gp p^+}
e^{{i \over 2}p^+\xmin +  {i \over 2}{m^2 \over p^+}
\xpl}. 
\label{F1}
\eea     
Using again the integral formulae \cite{GR}, 
we obtain the explicit expressions
\bea 
S_{22}(x) = &-&| {\ge}(\xpl)+{\ge}(\xmin)| \frac{m}{8}
\sqrt{\frac{\xpl}{\xmin}} \Big[J_1(m\sqrt{x^2}) + i~{\ge} (\xpl) N_1(m
\sqrt{x^2}) \Big] \nonumber \\ 
&-& \Big( {\ge}(\xpl) - {\ge}(\xmin) \Big)\frac{im}{4\gp}\sqrt{
\frac{\xpl}{\xmin}}K_1(m\sqrt{-x^2}) \nonumber \\
S_{11}(x) = &-&| {\ge}(\xpl)+{\ge}(\xmin)| \frac{m}{8}
\sqrt{\frac{\xmin}{\xpl}} \Big[J_1(m\sqrt{x^2}) + i~{\ge} (\xpl) N_1(m
\sqrt{x^2}) \Big] \nonumber \\ 
&+& \Big( {\ge}(\xpl) - {\ge}(\xmin) \Big)\frac{im}{4\gp}\sqrt{
\frac{\xmin}{\xpl}}K_1(m\sqrt{-x^2}) \nonumber \\
S_{12}(x) = &-&| {\ge}(\xpl)+{\ge}(\xmin)| \frac{m}{8}
\Big[N_0(m\sqrt{x^2}) - i~{\ge} (\xpl) J_0(m
\sqrt{x^2}) \Big] \nonumber \\ 
&+& | {\ge}(\xpl) - {\ge}(\xmin) |\frac{m}{4\gp}
K_0(m\sqrt{-x^2})
\label{fermicor}
\eea 
The corresponding expansions in the discretized case are  
\bea
\psi_2(x) = \frac{1}{\sqrt{2L}}\sum_{n=1}^{\infty}\Big[
b_n e^{-\frac{i}{2} p^+_n\xmin - \frac{i}{2}\frac{m^2}{p^+_n}\xpl} +
d^\dagger_n e^{\frac{i}{2} p^+_n\xmin + \frac{i}{2}\frac{m^2}{p^+_n}
\xpl}\Big] \nonumber \\
\psi_1(x) = \frac{m}{\sqrt{2L}}\sum_{n=1}^{\infty}\frac{1}{p^+_n}\Big[
b_n e^{-\frac{i}{2} p^+_n\xmin - \frac{i}{2}\frac{m^2}{p^+_n}\xpl} -  
d^\dagger_n e^{\frac{i}{2} p^+_n\xmin + \frac{i}{2}\frac{m^2}{p^+_n}
\xpl}\Big],
\label{dfermi}
\eea
where the fermion Fock operators satisfy
\beq
\big\{b_m,b^\dagger_n \big\} = \big\{d_m,d^\dagger_n \big\} = \gd_{mn}.
\label{ferfo}
\eeq 
To keep our discussion close to the scalar field case, we will work 
with periodic fermi field $\psi(-L)=\psi(L)$ (it is straightforward 
however to perform the whole analysis for antiperiodic fermi field).
The field equations (\ref{Dir}) again require zero modes of both 
components to vanish. 
 
The above field expansions yield for the correlation functions 
\bea
&&\hat{S}_{22}(x) = \frac{1}{2L}\sum_{n=1}^{\infty} e^{\frac{i}{2}
p^+_n\xmin + \frac{i}{2} \frac{m^2}{p^+_n}\xpl},\nonumber \\
&&\hat{S}_{11}(x) = \frac{m^2 L}{8 \gp^2}\sum_{n=1}^{\infty} \frac{1}{n^2}
e^{\frac{i}{2} p^+_n\xmin + \frac{i}{2} \frac{m^2}{p^+_n}\xpl},
\nonumber \\
&&\hat{S}_{12}(x) = \frac{m}{4\gp}\sum_{n=1}^{\infty} \frac{1}{n}e^{\frac{i}{2}
p^+_n\xmin + \frac{i}{2} \frac{m^2}{p^+_n}\xpl}.
\label{fermiw}
\eea

Do the discrete representations (\ref{df1}) and (\ref{fermiw}) lead to the 
continuum results (\ref{F1a}) and (\ref{fermicor}) for $L \rightarrow \infty$? 
More specifically, does 2 Im$\hat{D}$ reproduce the continuum Pauli-Jordan 
function in this limit? 
To answer this question, we begin by replacing the infinite series 
(\ref{df1}) and (\ref{fermiw}) by integrals using an integral representation of 
the polylogarithm functions.

\section{Integral representation of the discrete two-point function  
and the causality problem}

We start with the discretized two-point Wightman function (\ref{df1}). 
It is expressed 
as an infinite series. A very useful alternate representation of this function
can be given in the form of an integral based on polylogarithms. Consider the 
function of two independent complex variables defined by 
\bea
F_1(z,q) &=& \sum_{n=1}^{\infty}\frac{z^n} {n}e^{q/ n}.
\label{betaq}
\eea
For any finite $q$ it can be shown that the power series converges only within 
the unit circle $|z|<1$. Expanding $e^{q/n}$ in powers of its argument we 
obtain 
\beq
F_1(z,q)=\sum_{k=0}^{\infty}\frac{(q)^k}{k!}
\sum_{n=1}^{\infty} \frac{z^n}{n^{k+1}} =
\sum_{k=0}^{\infty}\frac{q^k}{k!}Li_{k+1} (z).
\label{rep1}
\eeq
Here we have used the definition \cite{Lew} of the polylogarithm function
$Li_m$,
\beq
Li_m(z)=\sum_{n=1}^{\infty}\frac{z^n}{n^m}.
\label{poldef}
\eeq
Note that this series representation (\ref{poldef}) of $Li_m$ converges only 
if $|z|<1$. Its analytic continuation to the rest of the complex $z$ plane is 
provided by the integral representation   
\beq
Li_m(z)=\frac{1}{(m-1)!}\int\limits_{0}^{\infty}du \frac{u^{m-1}}{z^{-1}
e^u-1}, (\;m\geq 1), 
\label{polyint}
\eeq
which shows that $Li_m(z)$ is actually analytic throughout the $z$-plane 
except for a cut on the positive real $z$ axis, linking $+1$ to $+\infty$.   
Substituting this formula into (\ref{rep1}) and interchanging integration
on $u$ with the summation on $k$, we arrive at the integral representation 
of the series (\ref{betaq}),    
\beq
F_1(z,q)=\int_{0}^{\infty}du\frac{1}{z^{-1}e^u-1}
I_0(2\sqrt{qu}).
\label{intf1}
\eeq
To obtain this result we have used the identity
\beq
\sum_{k=0}^{\infty}\frac{(v)^k}{(k!)^2}=I_0(2\sqrt{v}).
\label{I0}
\eeq
Comparing (\ref{df1}) and (\ref{betaq}) we note that 
\beq
\hat{D}(x)=\frac{1}{4\gp}F_1(e^{i\xi/Q},iQ),
\label{ourf}
\eeq 
where 
\beq
\xi=\gm^2\xpl\xmin/4,~~Q=\frac{\gm^2L}{4\gp}\xpl.
\label{variab}
\eeq
This is the first step of our analysis whereby the infinite series (\ref{df1}) 
has been rewritten in integral form via (\ref{intf1}) and (\ref{ourf}).  

There are four distinct cases to consider each associated with a quadrant of 
the $\xpl,\xmin$ plane. Consider first the case where both $x^{\pm}$ are 
positive ($Q>0, \xi>0$). We may rewrite Eqs.(\ref{intf1}) and (\ref{ourf}) as 
\beq
\hat{D}(\xpl>0,\xmin>0)=\frac{1}{4\gp Q}\int_{0}^{\infty}du\frac{1}
{\exp \Big(\frac{1}{Q}(u-i\xi)\Big) -1} I_0(2e^{i\gp/4}\sqrt{u}).
\label{plpl}
\eeq
The continuum limit is then obtained by considering the limiting value of 
the RHS of Eq.(\ref{plpl}) for $Q \rightarrow +\infty$. We now show that the 
result is the reduced version of Eq.(\ref{F1a}) appropriate to the regime 
$x^{\pm}>0$.  
Before proceeding we note that $I_0(2e^{i\gp/4}\sqrt{u})$ 
is an analytic function of $u$ throughout the entire finite complex plane
since its Taylor series expansion in powers of the argument has an infinite
radius of convergence. The remaining factor in the integrand of
(\ref{plpl}) is analytic throughout the $u$ plane with the exception of
simple poles at the discrete set of points $u_n=i(2n\gp Q+\xi)$ on the
imaginary axis, where $n$ is any integer. In view of these analytic properties
of the integrand we can alter the integration contour without changing the
value of the integral in (\ref{plpl}) as long as we avoid crossing through
any of the singular points $u_n$ and maintain the given endpoints. The simplest
choice of a preferable contour is presented in Fig.1. It consists of
the straight-line segments $(C_1)$ on the imaginary axis, the semi-circle
$(C_2)$ centered on the pole $u_0$, the line $(C_3)$ parallel 
to the positive real axis, and finally the straight-line segment $(C_4)$ 
parallel to the imaginary axis. Note that on $C_4$ we have ${\rm Re}~u = R$ 
and we will require that $R/Q>\ga Q$ where $\ga >2$.

On the contour $C_1$ we have $u=iv$, where $v$ is positive real. One
immediately finds that this contribution to $\hat{D}(x)$ may be written as
\beq
\frac{1}{8\gp Q} P \int\limits_{0}^{\gp Q+\xi}dv J_0(2\sqrt{v})\cot\big((v-\xi)
/(2Q)\big) - \frac{i}{8\gp Q}\int\limits_{0}^{\gp Q+\xi}dvJ_0(2\sqrt{v}), 
\label{Cone}
\eeq
where $P$ denotes principal value. Each of the two integrals is real and 
finite. The second term can be evaluated in closed form and the result is 
\beq
-\frac {i}{8\gp}\frac{\sqrt{\gp Q+ \xi}}{Q} J_1(2\sqrt{\gp Q+\xi}), 
\label{Conec}
\eeq
and in particular it vanishes in the large $Q$ limit.

In order to obtain the $L \rightarrow \infty$ limiting behavior of the first  
term of (\ref{Cone}), one can replace the cotangent function by the inverse 
of its argument, so that after the change of variable $v=w^2/4$ one obtains  
\beq
P \frac{1}{4\gp}\int\limits_{0}^{\infty}dv \frac{J_0(2\sqrt{v})}{v-\xi}=
P \frac{1}{2\gp}\int\limits _{0}^{\infty}dw\frac{wJ_0(w)}{w^2-4\xi} = 
-\frac{1}{4} N_0(2\sqrt{\xi}).
\label{f2}
\eeq
On the infinitesimal
semicircle $C_2$ defined by $u-i\xi=\ge e^{i\gt}, -\gp/2 \leq \gt
\leq \gp/2$, with $\ge \rightarrow 0^+$, we may use the approximation
\beq
\frac{1}{\exp\Big(\frac{1}{Q}(u-i\xi)\Big)-1} \approx \frac{Q}{\ge e^{i\gt}}.
\eeq
Replacing further the function $I_0$ by its value at $u_0$ and using the
relation \cite{GR} $I_0(2i\sqrt{\xi})=J_0(2\sqrt{\xi})$, we find that this
contribution to $\hat{D}$ is equal to $\frac{i}{16\gp}J_0(2\sqrt{\xi})$. 
Note that this quantity is independent of $L$ and thus survives in the 
large-$Q$ limit.

On the horizontal semi-infinite
line $C_3$ we may write $u=i\xi+ Q(i\gp + v)$, where $v$ is real, positive. 
The contribution to $\hat{D}(x)$ is given by 
\beq
-\frac{1}{4\gp}\int\limits_{0}^{\ga Q}dv\frac{1}{e^v+1}J_0\big(2\sqrt{\xi 
+ \gp Q - iQv}
\big). 
\label{Cthree}
\eeq
The asymptotic analysis of Eq.(\ref{Cthree}) in the limit $Q \rightarrow 
+\infty$ is very lengthy and involves substitution of an integral 
representation of $J_0$ combined with the method of steepest descent. The 
details of this calculation will be given elsewhere \cite{lmml}. The final 
result is that the leading behavior of the expression (\ref{Cthree}) is given 
by a term proportional to $\exp(iQ)$. This vanishes in the limit 
$L \rightarrow \infty$ as long as for any finite $L$ the quantity $x^+$ 
includes a small positive imaginary part such that $L\times {\rm Im}(x^+) 
\rightarrow \infty$. This requirement is satisfied for example by the choice 
Im$(x^+)=O(L^{-1/2})$.  

Finally, points on the line segment $C_4$ are described by $u=R+iv$, where 
$v$ is real, positive. The contribution to $\hat{D}(x)$ from $C_4$ is then 
found to be dominated by $\exp \Big[-\big(\ga - \sqrt{2\ga}\big)Q\Big]$ and 
thus vanishes in the large-$Q$ limit since we choose $\ga>2$. 

Combining these results we have
\beq
\lim _{Q \rightarrow \infty} \hat{D}(\xpl>0,\xmin>0) = -\frac{1}{4}
\Big(N_0(\gm\sqrt{x^2})-iJ_0(\gm\sqrt{x^2})\Big),
\label{fonefin}
\eeq
in agreement with Eq.(\ref{F1a}). 

It is easy to see from the formula (\ref{intf1}) that results for finite $L$ 
for the case $\xpl<0, \xmin<0$ ($\xi >0, Q<0)$ can be obtained from those  
for $\xpl>0,\xmin>0$ by complex conjugation:  
\bea
F_1(e^{-i\xi/|Q|},-i|Q|)=\big[F_1(e^{i\xi/Q},iQ)\big]^{*}.
\label{cc1}
\eea
Likewise, we have for $\xpl >0,\xmin<0$ ($Q>0,\xi <0$) 
\beq
\hat{D}(x)=\frac{1}{4\gp}F_1(e^{-i|\xi|/Q},iQ),
\label{q2}
\eeq
and this can be used to generate the results also for the regime $\xpl<0,
\xmin> 0$ $(\xi<0,Q<0)$ according to 
\beq
\hat{D}(x)=\frac{1}{4\gp}F_1(e^{i|\xi|/|Q|},-i|Q|) = 
\big[\frac{1}{4\gp}F_1(e^{-i|\xi|/Q},iQ)\big]^{*}. 
\label{q3}
\eeq
The evaluation of $F_1$ in Eq.(\ref{q2}) for large $L$ proceeds as above using 
the same multi-component contour except that the semicircle $C_2$ is not 
applicable since the pole $u_0$ is situated at $-i|\xi|$. The final result is 
\beq
\lim_{Q \rightarrow +\infty} \hat{D}(\xpl>0,\xmin<0) = \frac{1}{4 \gp}
\int\limits_{0}^{\infty}dv\frac{J_0(2\sqrt{v})}{v+\xi} = \frac{1}{2\gp}
K_0(2\sqrt{|\xi|}),
\label{imag1}
\eeq
in agreement with the continuum formula (\ref{F1a}). In particular, this 
result means that since the imaginary part of the function $\hat{D}(x)$ 
vanishes for spacelike $x^2$, the causality is restored in the 
infinite-volume limit. For large $L$ the leading $L$-dependent terms are 
of the order $L^{-1/4}$. 

Finally we briefly discuss the correlation functions $\hat{S}_{22}$ and 
$\hat{S}_{11}$ of (\ref{fermiw}). The first of these is a special case of the 
function
\beq 
F_0(z,q) = \sum_{n=1}^{\infty}{z^n}e^{q/ n} = \frac{z}{1-z}  
+q \int\limits_{0}^{\infty}du\frac{1}{z^{-1}e^u-1}\frac{I_1(2\sqrt{qu})}
{\sqrt{qu}}, 
\label{intf22}
\eeq
while $\hat{S}_{11}$ is a special case of 
\beq 
F_2(z,q) = \sum_{n=1}^{\infty}\frac{z^n}{n^2}e^{q/ n} =   
\int\limits_{0}^{\infty}du\frac{u}{z^{-1}e^u-1}\frac{I_1(2\sqrt{qu})}
{\sqrt{qu}}. 
\label{intf11}
\eeq
In the regime $x^{\pm}>0$ we find that
\bea
\lim_{Q \rightarrow +\infty} \hat{S}_{22}(x) &=& 4e^{i\frac{\gp}{4}}P 
\int\limits_{0}^{\infty}dw\frac{J_1(w)}{w^2-4\xi} - \frac{\gp}{\sqrt{\xi}}
e^{-i\frac{\gp}{4}}J_1(2\sqrt{\xi}) \nonumber \\
&=& -\frac{m}{4}\sqrt{\frac{\xpl}{\xmin}} 
\Big(J_1(m\sqrt{x^2}) + i N_1(m\sqrt{x^2})\Big).
\label{s22}
\eea
Likewise, in the same regime, we have 
\beq
\lim_{Q \rightarrow +\infty} \hat{S}_{11}(x) =  
-\frac{2\gp^2}{m}\sqrt{\frac{\xmin}{\xpl}} 
\Big(J_1(m\sqrt{x^2}) + i N_1(m\sqrt{x^2})\Big).
\label{s11}
\eeq
These results are in agreement with the continuum formulas (\ref{fermicor}).
 
\section{Numerical results}

In principle, one may try to evaluate the integral (\ref{intf1}) representing 
the Pauli-Jordan function in finite volume numerically for increasing values 
of $L$ to examine the rate of convergence towards the continuum result. 
In practice, this is rather difficult to achieve since the integrand 
of the representation (\ref{intf1}) oscillates rapidly due 
to the presence of the Bessel function $I_0$. Already for relatively 
small values of $Q$ the amplitudes of these oscillations are so large and 
the spacings of successive zeros are so small that it is very difficult to 
reliably evaluate the integral by standard numerical routines. Since it is not 
our goal to perform extensive numerical analyses in the present work, we have 
computed the integral for a few relatively small values of $Q(=4,15,18)$ using  
an integration method based on Chebyshev polynomials as well as by a 
Clenshaw-Curtis method. For definiteness we set $\mu^2 \xpl=1$ so that the 
corresponding box lengths given by $L=4\pi Q$ are approximately $L=50, 188$ 
and $226$. The results are displayed in Fig.2. An essential difference in 
the behavior of the Pauli-Jordan commutator function between the space-like 
region (negative values of $\xmin/L$) and time-like region (positive 
$\xmin/L$) is obvious already for the smallest value $Q=4$. It is also evident 
that for larger Q the oscillatory behavior of the continuum curve in the 
time-like region is resolved with increased accuracy. This is particularly true 
in the interval $0<\xmin/L<0.5$ but the number and position of oscillations is 
semiquantitatively reproduced also for $\xmin/L > 0.5$. Although the 
Pauli-Jordan function for finite volumes is not zero in the space-like region, 
it is reasonably close to it. We recall that for our choice of $Q$ values we 
are still very far from the infinite-volume limit so the obtained behavior of 
the Pauli-Jordan function is very plausible and consistent with our analytical 
results. 

\section{Discussion}

The Discretized Light Cone Quantization method has led to many interesting
results over the last fifteen years. It is however essential that the method 
satisfy all requirements and principles of a consistent relativistic theory. 
It is rather clear that restricting a system to a finite spatial volume and 
imposing (anti)periodic boundary conditions on quantum fields will generally 
lead to a theory which does not have exactly the same properties as the 
continuum (infinite-volume) theory. Some finite-volume effects may be present
\cite{Haag}. However, the crucial point is that the continuum limit of a
finite-volume theory should recover all necessary properties including e.g. 
Poincar\'e symmetry. The principle of microcausality (or locality) is one of 
the fundamental properties of relativistic dynamics and the DLCQ method would 
face a serious difficulty if it would be in conflict with this principle. We 
have demonstrated in the present work analytically as well as numerically that 
this is not the case.  With the infinite number of field modes the violation 
of microcausality in a LF finite volume with periodic scalar field is only a 
marginal effect and continuum results including the causal behavior are 
restored in the $L \rightarrow \infty$ limit. In practice, the DLCQ 
calculations of mass spectra and wavefunctions are always performed for finite 
$L$ and with a finite number of Fouri\'er modes. At this step, the causality 
may seem to be violated \cite{HKS} (see also \cite{Haricau} for a treatment 
that averages over some range of $L$ values and restores the causal behavior 
in a finite volume). However, as physical quantities calculated with the DLCQ 
method have to be extrapolated to the continuum limit, there is no 
inconsistency, since, as we have shown, the causality is restored there. 
 
Our conclusion has been obtained by means of a well defined mathematical 
treatment leading to an integral representation of the infinite series 
representing the two-point functions for a finite volume. We have extracted the 
$L$-independent part of the integral as well as the leading large-$L$ 
corrections. Our results for the PJ function $2{\rm Im}\hat{D}(x)$ are 
consistent with the results of Ref.\cite{GSW}. The method used enabled us 
however to calculate the complete Wightman functions of two-dimensional free 
massive bosons and fermions quantized in a finite volume. A detailed 
discussion of our mathematical treatment will be published separately 
\cite{lmml}.

Ames Laboratory is operated for the United States Department of Energy by Iowa 
State University under Contract No. W-7405-Eng-82. One of the authors (L.M.) 
was partially supported by the Slovak Grant Agency, VEGA Grant No.2/3106/2003. 

\section{Acknowledgements}

The authors would like to thank J. P. Vary for support and Peter Marko\v s 
for assistence with the numerical calculations.

\begin{figure}
\includegraphics[width=5.10in]{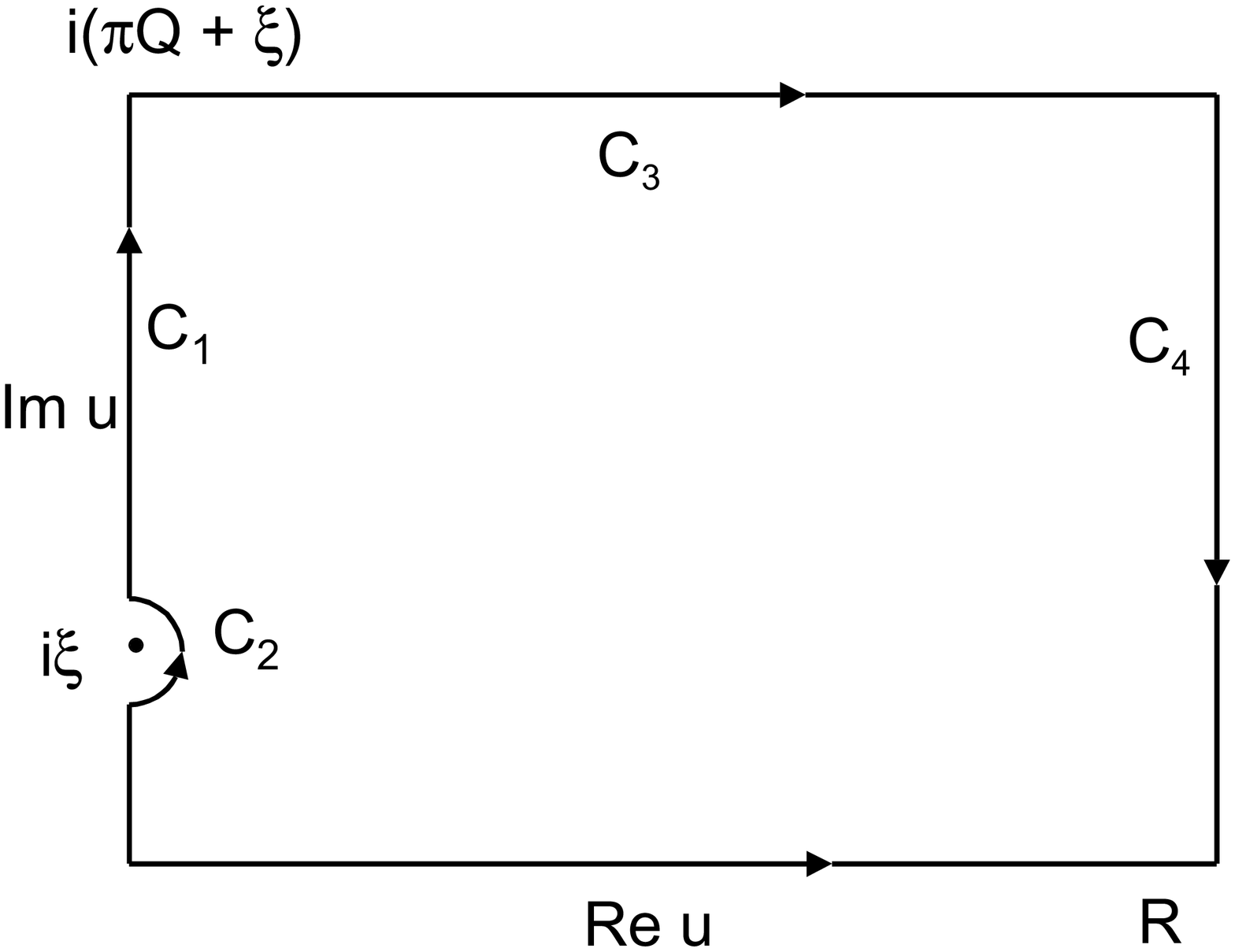}
\caption{Integration contour chosen for our evaluation of the integral 
(\ref{intf1}).}
\label{fig1}
\end{figure} 

\begin{figure}
\begin{center}
\psfig{figure=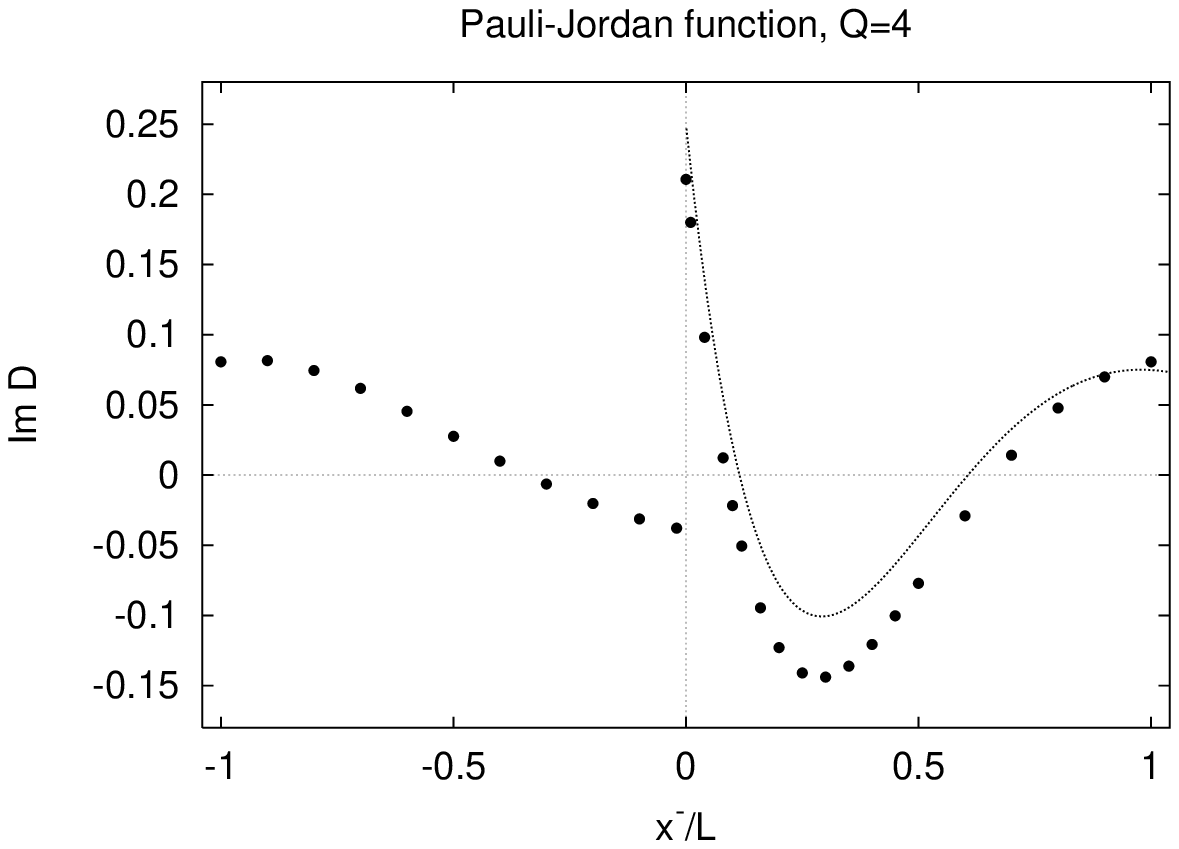,height=52mm,width=100mm}
\psfig{figure=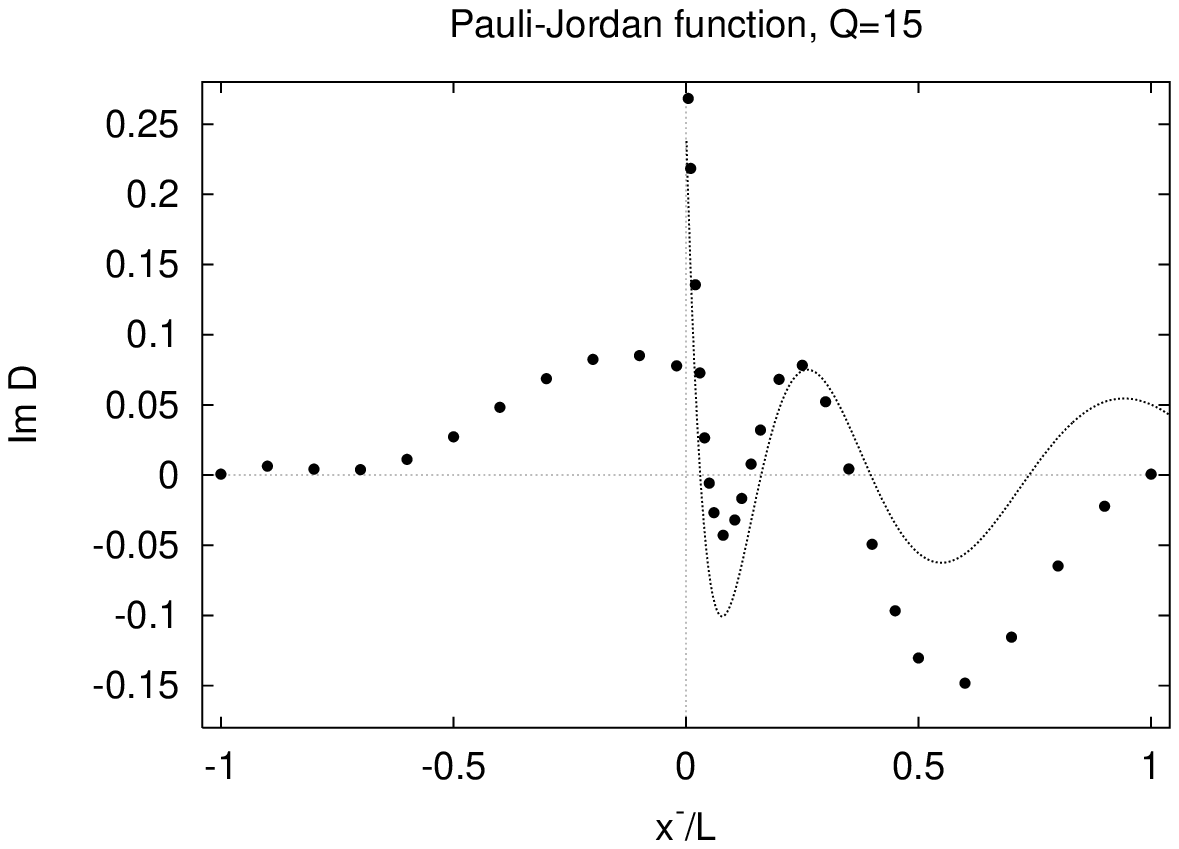,height=52mm,width=100mm}
\psfig{figure=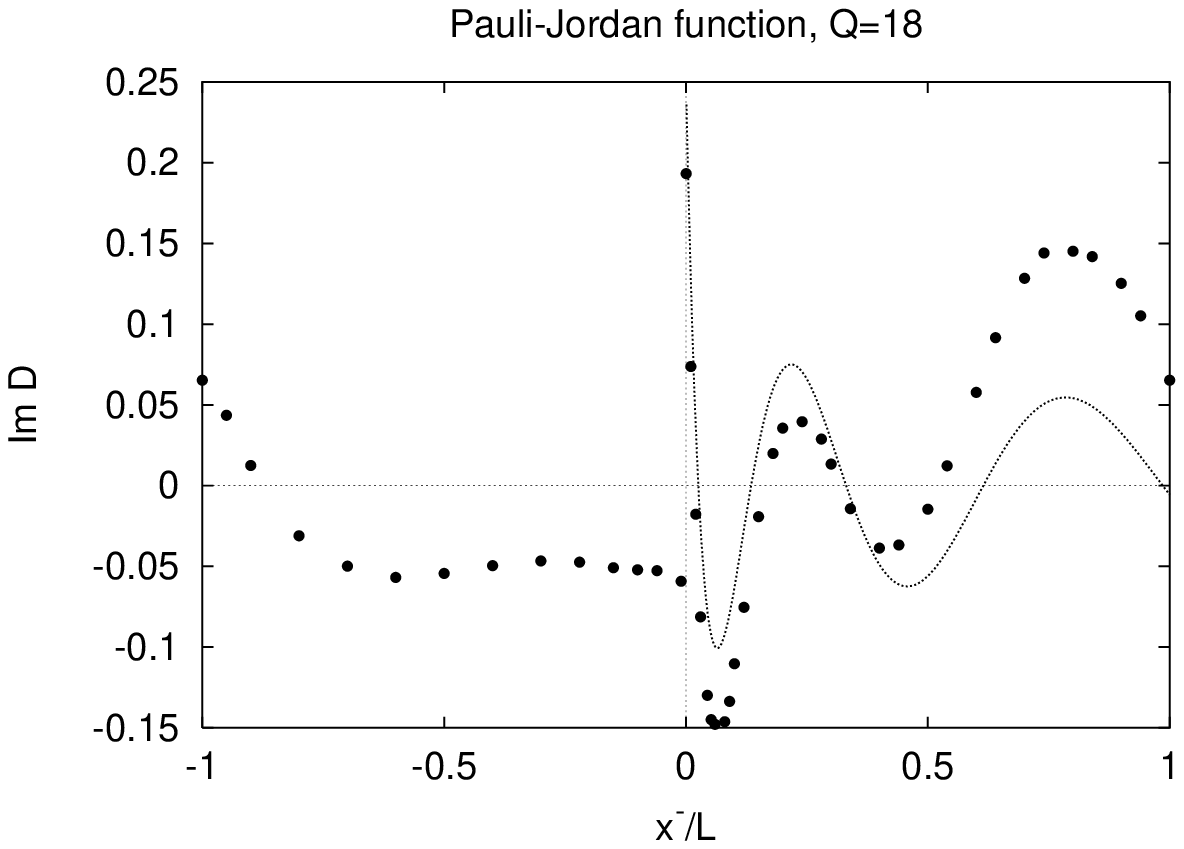,height=52mm,width=100mm}
\caption{The commutator function ${\rm Im} D(\xpl,\xmin)$ evaluated numerically 
is plotted for three values of $Q$ as a function of $s=\xmin/L$ and compared 
with the continuum function $1/4 J_0(\sqrt{Ls})$ (solid line) in the 
box of the length $L=4\pi Q$. For simplicity, we chose 
$\mu^2 \xpl =1$. The time-like region corresponds to positive values of $s$, 
the space-like region to negative ones.}
\label{fig2}
\end{center}
\end{figure} 

\end{document}